\documentclass{iopart}
\usepackage[pctex32]{graphicx}

\begin{document}

\title{Detrended fluctuation analysis as a statistical tool to 
monitor the climate }

\author{M.L. Kurnaz}
\address{Department of Physics, Bogazici University, 34342
Bebek Istanbul Turkey}
\ead{kurnaz@boun.edu.tr}

\begin{abstract}

Detrended fluctuation analysis is used to investigate power law 
relationship between the monthly averages of the maximum daily 
temperatures for different locations in the western US. On the 
map created by the power law exponents, we can distinguish different 
geographical regions with different power law exponents. When the 
power law exponents obtained from the detrended fluctuation analysis 
are plotted versus the standard deviation of the temperature 
fluctuations, we observe different data points belonging to the 
different climates, hence indicating that by observing the long-time 
trends in the fluctuations of temperature we can distinguish between 
different climates.
\end{abstract}

%\maketitle

\section{INTRODUCTION\protect\\ }
\label{sec:level1}

Recent advances has shown us that we can use scaling arguments to 
analyze climatic data \cite{Bunde100, Kantelhardt13100, Koscielny-Bunde380,
Eichner13120, Livina13130, Monetti13140, Bunde13180, Govindan13190, 
Vjushin13200, Kantelhardt13220, Govindan13230, Koscielny-Bunde13240, 
Koscielny-Bunde13250, Bodri13500, Bodri13530, Weber2830, Talkner3550, 
Weber6080, Fraedrich13270}. The ongoing discussion focuses on whether 
we can obtain a correlation between the scaling exponents obtained 
from these analyses and the geographic location of the weather stations 
\cite{Bunde100, Eichner13120, Fraedrich13270, Bunde12710, Fraedrich12720}.

Determining the weather is a rather simple issue. A cold day is 
usually followed by a cold day, and a warm day is usually followed 
by a warm day. On a larger scale, a colder week is usually followed 
by a warmer week which corresponds to the average duration of the 
general weather regimes. But as the longer timescales are governed 
by different processes like the circulation patterns and sometimes 
even influenced by trends like global warming, defining long-term 
correlations becomes more difficult. 

In order to separate the trends and the correlations we need to 
eliminate the trends in our temperature data. Several methods are 
used effectively for this purpose: rescaled range analysis (R/S) 
\cite{Bodri13500, Bodri13530, Mandelbrot12180, Mandelbrot12190}, 
wavelet techniques (WT) \cite{Koscielny-Bunde13240, Arneodo5250} 
and detrended fluctuation analysis (DFA) \cite{Koscielny-Bunde13250, 
Peng13540}.

Analysis of the temperature fluctuations over a period of decades in 
different places of the globe has already showed the effectiveness of 
the application of detrended fluctuation analysis to characterize the 
persistence of weather and climate regimes. DFA and WT have been applied 
to study temperature and precipitation correlations in different climatic 
zone on the continents and also in the sea surface temperature of the 
oceans. The recent results show that the temperatures are long range power 
law correlated. The long-term persistence of the temperatures can be 
characterized by an auto-correlation function $C(n)$ of temperature 
variations where $n$ is the time between the observations. 
The auto-correlation function decays as $C(n)\sim n^{-\gamma}$. Even though there is some 
disagreement on the value of the exponent $\gamma$, the fact that the persistence 
of the temperatures can be characterized by this auto-correlation function 
is firmly established. Different groups have used R/S, DFA and WT analysis 
and have shown that this exponent $\gamma$ has roughly the same value 
$\gamma \simeq 0.7$  for the continental stations \cite{Eichner13120, Bunde13180, 
Kantelhardt13220, Koscielny-Bunde13250, Bodri13500, Bodri13530, Weber2830, 
Talkner3550}.  The exponent $\gamma$ is found to be roughly 0.4 for 
island stations \cite{Eichner13120, Bunde13180} and sea surface temperature 
on the oceans \cite{Monetti13140, Bunde13180}. This method has also been 
applied to the temperature predictions of coupled atmosphere-ocean general 
circulation models \cite{Bunde100, Govindan13190, Fraedrich13270, Fraedrich12720, 
Blender13260} but there is disagreement on the actual value of the exponent 
$\gamma$. On one side it is argued that the exponent does not change with the 
distance from the oceans \cite{Bunde100, Eichner13120} and is roughly 
$\gamma \simeq 0.7$. On the other side it is said that the scaling exponent 
is roughly 1 over the oceans, roughly 0.5 over the inner continents and about 
0.65 in transition regions \cite{Fraedrich12720}.

Previous work in this area also shows that there is a slight variation in the 
scaling exponent between the low-elevation, mountain, continental and maritime 
stations \cite{Weber2830, Talkner3550}. Even though these variations are 
between  $\gamma \simeq 0.5$ and $\gamma \simeq 0.7$ the fact that they show a 
correlation with location and elevation indicates that a relationship between 
the statistical nature of the temperature fluctuations and the climate can be 
established.

The work presented in this paper focuses on the statistical aspect of the 
temperature data and tries to establish a method to analyze temperature 
records in such a way as to reveal long term trends in the climate with the 
hope to help the models created to measure climate change.

\section{METHOD\protect\\ }
\label{sec:level2} To remove the seasonal trends that are known to exist 
in the daily temperature data we need to determine the mean temperature 
for each day over all the years in the time series. We then calculate the 
fluctuation of the daily temperature from the mean daily temperature,
\begin{equation}
\Delta T_{i} = T_{i} - <T_{i}>
\end{equation}
where $<T_{i}>$ is the mean daily maximum temperature. Similarly we can 
also use the mean daily average temperature or the mean daily minimum 
temperature, and the use of average or minimum temperature instead of 
maximum temperatures does not change the outcome of the analysis 
\cite{Talkner3550}. To remove the remaining linear trends in the data 
(the average temperature for some years can be higher or lower than the 
average temperature of the time series for that location as a result of 
long-term atmospheric processes), we applied Detrended Fluctuation 
Analysis (DFA) method \cite{Peng13540}, which is used to remove trends 
in the data to allow investigation of long-term correlations in the data.

The noise and the nonstationarity in the temperature data usually hinders 
a reliable and direct calculation of the autocorrelation function $C(n)$, 
instead we calculate a running sum of the temperature fluctuations,
\begin{equation}
y (m) = \sum_{i=1}^n \Delta T_{i}
\end{equation}
where $m = 1,. . . , n$. Next, the time series of the $y(m)$ is divided into 
nonoverlapping intervals of equal length $n$. In each interval, we fit 
$y(m)$ to a straight line, $x(m) = km + d$ for each segment and calculate 
the detrended square variability $F^{2}(n)$ as
\begin{equation}
F^{2}(n) = < \frac{1}{n} \sum_{m = kn + 1}^{(k + 1)n} (y(m) - x(m))^{2}>
\end{equation}
with
\begin{equation}
k = 0, 1, 2, . . . , ( \frac{N}{n} - 1).
\end{equation}

If the temperature fluctuations were uncorrelated (white noise) we would expect
\begin{equation}
F(n) \sim n^\alpha
\end{equation}
where  $\alpha = \frac{1}{2}$. If $\alpha > \frac{1}{2}$, we would expect 
long-range power law correlations in the data for the range of values 
considered. Figure 1 shows an example of such an analysis method for a sample 
station, Cheyenne WSFO, WY. The data follows a straight line with slope 
$\alpha = 0.58 \pm 0.01$. Even though there is some scatter in the data after 
a period of 10 years, the scatter in the data is still within the error bars 
of the analysis. This scatter is caused by the fact that the average in the 
detrended square variability $F^{2}(n)$ has been taken over larger and larger 
values of $n$, resulting in poor statistics. The dataset consists of 107 years 
of monthly temperature data between the years 1888-1994.

\begin{figure}
\begin{center}
\includegraphics[width=0.80\textwidth]{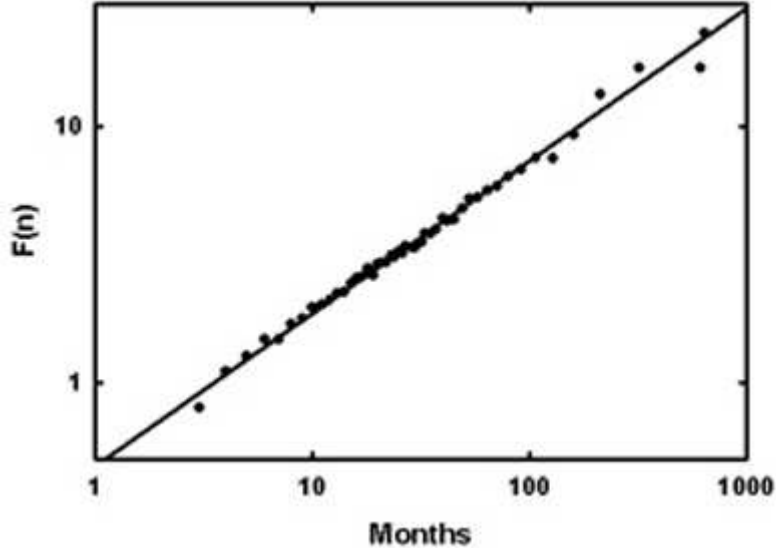}
\caption{\label{Fig1}The power law relationship between the time period 
and the detrended square variability. The data is for a sample station, 
Cheyenne WSFO, WY with slope $\alpha = 0.63 \pm 0.01$.}
\end{center}
\end{figure}

The main problem when we want to adopt this analysis to climate is the absence 
of high quality long-term daily temperature data. When we need to look at the 
short term fluctuations in daily temperatures, we must obtain reliable data, 
however if our aim is long-term behavior of temperature data we can also use 
monthly averages instead of daily averages. It has been shown that for the 
time domain longer than 60 days, both the daily and monthly temperature data 
give us the same scaling exponent \cite{Weber2830, Talkner3550}. As our 
analysis for daily and monthly data at all the sampled locations also agrees 
for long time scales ($n > 60 days$) the results presented in the paper have 
been obtained from the analysis of monthly averages. 

In the present work we have investigated temperature fluctuations for 384 
weather stations in the Western US. The data has been obtained from the 
U.S. Historical Climatology Network \cite{US13720}. From the available data 
we have chosen the stations with the longest records. We did not include in 
our analysis datasets with data shorter than 75 years, and the longest 
dataset we had was 110 years at several locations in the dataset (all 
ending in 1994). 

Previously it has been observed that power law exponents obtained from DFA 
of temperature fluctuations stay between 0.55 and 0.70 \cite{Bunde100, 
Eichner13120, Govindan13190, Koscielny-Bunde13240, Koscielny-Bunde13250}. 
Figure 2 gives a summary of the scaling exponents obtained from the 384 
stations in the Western US. As we can see from this figure, consistent 
with the earlier observations, we obtain scaling exponents in the range 
of 0.50 to 0.74. The average for all the dataset investigated comes out to be 
$\bar{\alpha} = 0.61 \pm 0.04$.

\begin{figure}
\begin{center}
\includegraphics[width=0.80\textwidth]{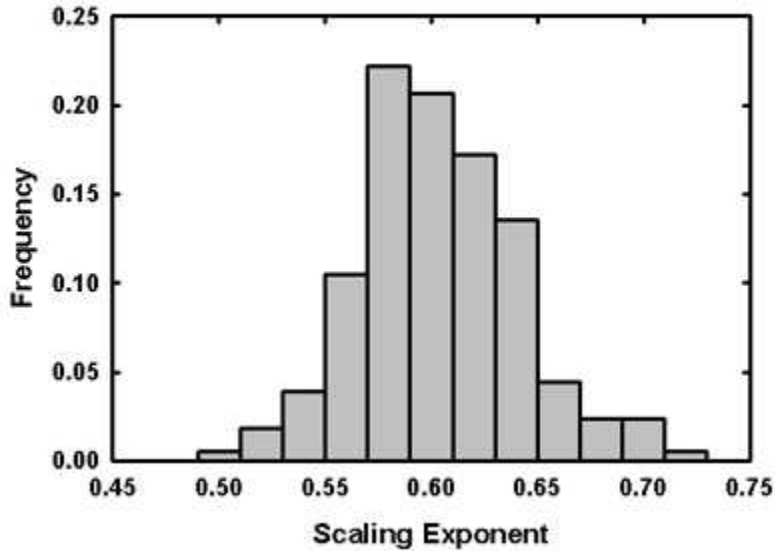}
\caption{\label{Fig2}The histogram of the scaling exponents is shown. 
The average from 384 stations gives a scaling exponent of 
$\bar{\alpha} = 0.61 \pm 0.04$.}
\end{center}
\end{figure}

Figure 3 shows a map of all the power law exponents for all the stations. 
This map has been generated using the data using a running average method 
to generate a grid of 30 x 30 regions with the data we have. A simple 
observation about the power law exponents is even though the power law 
exponents crowd around the mean value, their special distribution is not 
uniform. From the map we can simply distinguish a few regions: Western part 
of Washington and Oregon, Montana region, Inland California, Utah and New 
Mexico. Table 1 gives the power law exponents for these regions. This table 
is sufficient to divide the western US into two different regions, one with 
high power law exponent $\bar{\alpha} = 0.64 \pm 0.04$ and one with lower 
power law exponent $\bar{\alpha} = 0.59 \pm 0.01$.

\begin{figure}
\begin{center}
\includegraphics[width=0.80\textwidth]{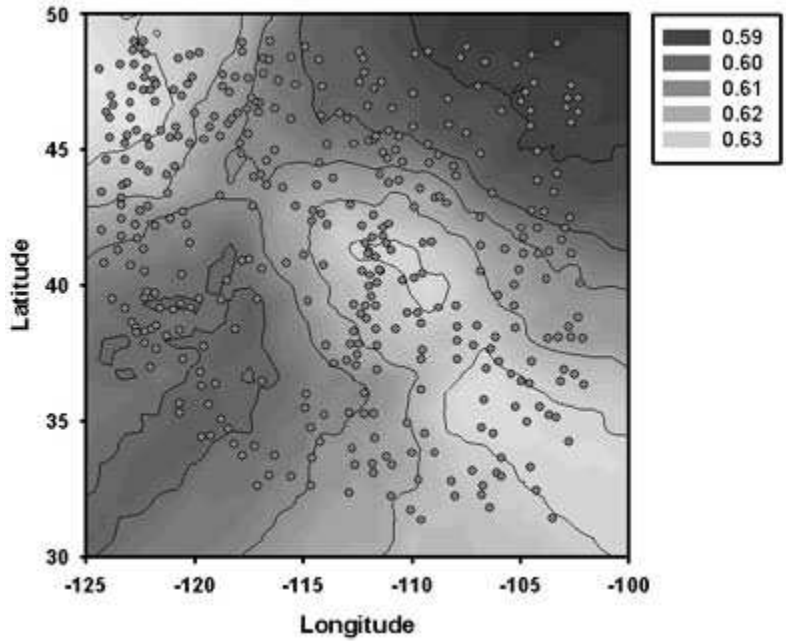}
\caption{\label{Fig3}The map of the power law exponents for the western 
US. The stations are presented by circles in the data. The interpolation 
has been done using the running average method.}
\end{center}
\end{figure}

However, such a division is not sufficient because of two main reasons:
\begin{itemize}
\item	Western Oregon and New Mexico are both categorized as high power law 
exponent regions. But when we look at the climate of these regions we see 
very different characteristics, western Oregon stations predominantly 
being Humid Subtropical Mediterranean ($83 \%$ of the stations in this 
region) whereas New Mexico is mostly dry/arid (cool) mid latitude 
desert ($88 \%$ of the stations in this region).
\item New Mexico and Coastal California are categorized as high and low 
power law exponent regions respectively. But when we look at the climate 
of these regions we see very similar characteristics, both regional 
stations mostly being dry/arid (cool) mid latitude desert (New Mexico - 
$88 \%$, coastal California - $70 \%$ of the stations).
\end{itemize}

However a striking result can be observed if we plot the standard 
deviation of the temperature fluctuations versus the power law exponent 
observed from these stations. Figure 4 gives a summary of the power law 
exponents for the six regions mentioned in Table 1, Western Washington 
and Oregon, New Mexico, Utah, Coastal California, Inland California and 
Western Nevada, and Montana, versus the standard deviation of the 
temperature fluctuations, respectively. In this figure we can clearly 
identify that, with the aid of the standard deviation of temperature 
fluctuations, the scaling exponents can be used to distinguish between 
different climates, and perhaps bring some refinements to the already 
existing classification \cite{Koeppen13800}. 

\begin{figure}
\begin{center}
\includegraphics[width=0.80\textwidth]{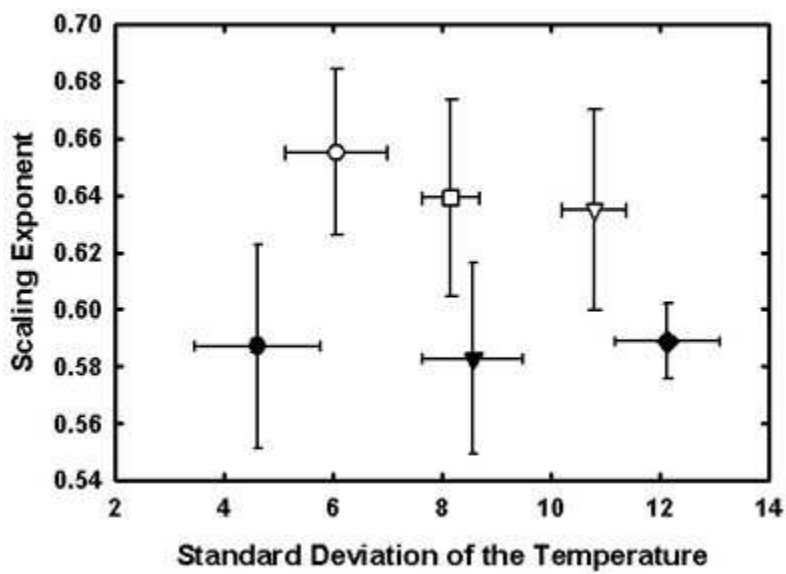}
\caption{\label{Fig4}The scaling exponents plotted against the standard 
deviation of the temperature fluctuations for Western Washington and 
Oregon (empty circles), New Mexico (empty squares), Utah (empty triangles), 
Coastal California (full circles), Inland California and Western Nevada 
(full triangles), and Montana (full diamonds).}
\end{center}
\end{figure}

\begin{center}
\begin{table}
\caption{Power law exponents for different regions of Western United States.} 
\label{sums}
\begin{tabular}{|c|c|}
\hline Region: & Power Law Exponent: \\ 
\hline Western Oregon and Washington & $0.66 \pm 0.03$ \\ 
\hline Utah and Western Colorado & $0.64 \pm 0.03$ \\ 
\hline New Mexico, Oklahoma panhandle, Southern Colorado & $0.64 \pm 0.03$ \\ 
\hline Inland California and Western Nevada & $0.58 \pm 0.03$ \\ 
\hline Montana and Western North Dakota & $0.59 \pm 0.01$ \\ 
\hline Coastal California & $0.59 \pm 0.03$ \\ \hline
\end{tabular}%
\end{table}
\end{center}

\section{DISCUSSION\protect\\ }
\label{sec:level3} We have investigated the power law behavior of the 
temperature fluctuations to gain more insight to a statistical 
description of different climate types. The main question in this 
regard that would come to mind is why not simply use the regular way of 
measuring temperature, pan evaporation and precipitation to determine 
the climate type for a region. Most certainly we do not need a sharper 
method than using the meteorological data for the present climate in 
our world, however for two major reasons we would need a statistical 
method to describe the climates. 

When we look at the climatic data and how it has changed in the past, 
we need a method to analyze paleoclimatic data to reveal structure of 
timescales not only of the order of decades but millions of years. 
But the difficulty shows itself as the paleoclimatic data come from 
a number of different data sources and they have different units. But 
only one thing is common to all of this data, and that is the fact that 
all of this data is in terms of a time series of amplitudes. Our results 
suggest that on the basis of power law scaling of the temperature 
fluctuations of the maximum daily temperatures we can distinguish between 
different climates. The real challenge comes when we try expand this 
analysis into the past, as to our knowledge, no reliable monthly data 
exists beyond 218 years \cite{Koscielny-Bunde13240}. However, previous 
work where authors have established long-range power law behavior using 
rescaled-range analysis \cite{Fluegeman9510, Wang7920} gives us hope 
about expanding the use of the method we have developed.

When we use global climate models to extrapolate the climate changes into 
the future, the method of detrended fluctuation analysis proves to be a 
powerful tool \cite{Govindan13190, Fraedrich13270}. When the length scale 
and the future time scale of application of these climate models improves 
the present work may help to fill in the voids in the models by providing 
information on how the climate would be like.

\ack
Data were provided by U.S. Historical Climatology Network and the National 
Climatic Data Center.

\end{document}